\input harvmac
\def \T {{\rm T}}
\def \FF {\hat F}
\def \F {{\cal F}}
\def \tr {{\rm tr}}
\def \gym {g_{_{\rm YM}}}

\def \a {\alpha}

\def \ep{\epsilon}

\def \N {{\cal N}}

\def \x {\xi} 
\def \C {{\cal C}}

\def \del {\partial}

\def \const {{\rm const}}

\def \b {\beta}

\def \s {\sigma}
\def \p {\phi}
\def \m {\mu}
\def \n {\nu}

\def \td {\tilde }
\def \d {\delta}

\def \C {{\cal C}}
\def \diag {{\rm diag}}
\def \P {\Phi}

\def \inv {^{-1}}
\def \ov {\over }

\def \fourth{{{1\over 4}}}

\def \ha {{ {1 \over 2}}}

\def \ni {{K}} 
\def \hi {H_{-1}} 
\def \pa{+}

\def \l {\lambda}

\def \x {\xi} 
\def \C {{\cal C}}
\def \e {{\eta}}

\def \T {{\cal T}} 
\def \lr { \lref}
\def\np {{  Nucl. Phys. }}
\def \pl {{  Phys. Lett. }}

\def \pr  {{ Phys. Rev. }}


\lr\malr{ J. Maldacena, { Wilson Loops in Large N Field Theories,}
Phys. Rev. Lett. { 80} (1998) 4859, hep-th/9803002;
S-J. Rey and J. Yee, { Macroscopic Strings as heavy quarks
in Large N Gauge Theory,} { hep-th/9803001}. }

\lr\itzh{ N. Itzhaki, J. Maldacena, C. Sonnenschein  and  S. Yankielowicz,
Supergravity and the Large N Limit of Theories with 16 Supercharges,
{ Phys. Rev.} { D58} (1998) 046004,
{ hep-th/9802042.}}

\lr\dine {  M.  Dine   and N.  Seiberg,  Comments  on higher derivative operators
in some susy field theories, 
    Phys. Lett. B409  (1997) 239,  
    hep-th/9705057;
S.  Paban, S. Sethi  and M.  Stern, Supersymmetry and higher derivative
 terms in  the effective action of Yang-Mills theories, 
        JHEP 06  (1998) 012, 
    hep-th/9806028;     
 F. Gonzalez-Rey, B. Kulik, I.Y. Park and  M. Ro\v cek,
Selfdual effective action of N=4 super Yang-Mills,
hep-th/9810152; E.I. Buchbinder, I.L. Buchbinder  and S.M. Kuzenko, 
Nonholomorphic effective potential in N=4 SU(n) SYM,
Phys. Lett. B446 (1999) 216,  hep-th/9810239;
 D.A. Lowe and  R. von Unge, 
Constraints on higher derivative operators in maximally supersymmetric gauge theory,   JHEP 9811 (1998) 014, hep-th/9811017;
S. Hyun, Y. Kiem and  H. Shin,
Supersymmetric completion of supersymmetric quantum mechanics,
 hep-th/9903022. 
}

\lr \dougtay{ 
M.R. Douglas and  W. Taylor,
Branes in the bulk of Anti-de Sitter space,
hep-th/9807225.}

\lr\kall{R. Kallosh  and  A.A. Tseytlin,
Simplifying superstring action on  $AdS_5 \times S^5$,  
 JHEP 9810  (1998) 016,
 hep-th/9808088.}

\lr \dkps{  M.R. Douglas, D. Kabat, P. Pouliot and S.H. Shenker,
 \np B485 (1997) 85,
hep-th/9608024. }

\lr \mik {L. Girardello, M. Petrini, M. Porrati and  A. Zaffaroni,
Confinement and condensates without fine tuning in supergravity duals of gauge theories, 
 hep-th/9903026.
 }

\lr \min { J. Minahan, 
Asymptotic freedom and confinement from type 0 string theory, 
hep-th/9902074.
 }

\lr \koc{R. de Mello Koch, A. Paulin-Cambell and J.P. Rodrigues, 
Non-holomorphic corrections from threebranes in F theory, hep-th/9903029.}

\lr\igo{ I.R. Klebanov, World volume approach to absorption by nondilatonic branes,
  {Nucl. Phys.} {B496} (1997) 231,
  hep-th/9702076.
}

\lr\nnn{
A. Kehagias, The new type IIB vacua and their F-theory interpretation, 
\pl B435 (1998) 337, hep-th/9805131;
O.  Aharony, A. Fayyazuddin and J. Maldacena, 
The large N limit of N=2,1 field theories from threebranes in F-theory, 
JHEP 9807 (1998) 013, hep-th/9806159;
C. Ahn, K. Oh  and  R. Tatar, 
The Large N limit of N=1 field theories from F theory,
hep-th/9808143;
R. de Mello Koch and R. Tatar, Higher derivative terms from three-branes in F theory, hep-th/9811128.
}

\lr\gub {S.S. Gubser, { Dilaton-driven confinement,} { hep-th/9902155.}}

\lr \kes {A. Kehagias and K. Sfetsos, { On running couplings in gauge
theories from IIB supergravity,} { hep-th/9902125}}

\lr\hord {G.T. Horowitz and A.~Strominger, Black strings and p-branes,  Nucl.
  Phys. B360 (1991) 197; 
 M.J. Duff  and J.X. Lu, 
The self-dual  type IIB  superthreebrane,
{Phys. Lett.}  {B273} (1991)  409.}

\lr\gik{ 
G.W. Gibbons, M.B. Green and  M.J. Perry,
Instantons and seven-branes in type IIB superstring theory,
 Phys. Lett. B370 (1996) 37, 
 hep-th/9511080. 
 }

\lr\gross{D.J. Gross and  H. Ooguri, Aspects of large N gauge theory 
dynamics as seen by string theory, 
Phys. Rev. D58 (1998) 106002, 
hep-th/9805129.} 

\lr\tsee{ A.A. Tseytlin, Harmonic superpositions of M-branes,
\np B475 (1996) 149,  hep-th/9604035;
Type IIB instanton as a wave in twelve-dimensions,
 Phys. Rev. Lett.  78 (1997) 1864, hep-th/9612164.}

\lr \mink{ P. Minkowski, On the ground state expectation values of the field strength bilinears in gauge theory and constant classical fields, 
\np B177 (1981) 203.}

\lr \leit{H. Leutwyler, Constant gauge fields and their quantum fluctuations, \np B179 (1981) 129.}

\lr \oth  {C.A. Flory, A self-dual gauge field, its quantum fluctuations,
and interacting fermions, \pr D28 (1983) 3116; 
P. van Baal, SU(N) Yang-Mills solutions with constant 
field strength on $T^4$, 
Commun. Math. Phys. 94 (1984) 397;
Yu.A. Simonov, Vacuum background fields in QCD as a source of confinement, 
\np B307 (1988) 512;
 Yu.A. Simonov, Confinement, Phys. Usp. 39 (1996) 313,  hep-ph/9709344.
 }

\lr\kli{K. Lee,  Sheets of BPS Monopoles and Instantons with Arbitrary Simple Gauge Group, hep-th/9810110.}

\lr \ft{S.J. Gates, M.T. Grisaru, M. Ro\v cek  and  W. Siegel,
    Superspace  
    (Benjamin/Cummings, 1983); 
E.S. Fradkin and A.A. Tseytlin,
Quantum properties of higher dimensional and dimensionally 
reduced supersymmetric theories, Nucl. Phys. B227 (1983) 252. }

\lr \dougli {  M.R. Douglas and  M. Li, 
D-brane realization of $N=2$ 
 super Yang-Mills theory  in four dimensions,
hep-th/9604041.}
\lr\mmm{
 J. Maldacena, Probing near-extremal black holes with D-branes,  Phys. Rev. D57  (1998) 3736,  hep-th/9705053.}
\lr\chee{I. Chepelev and  A.A. Tseytlin,  Long-distance interactions of branes: correspondence between supergravity and super Yang-Mills
      descriptions, 
 Nucl. Phys. B515 (1998) 73, hep-th/9709087.
}

\lr\chept{
  I. Chepelev  and  A.A. Tseytlin,
Interactions of type IIB D-branes from D-instanton matrix model,
 Nucl. Phys. B511 (1998) 629, hep-th/9705120.}

\lr \mald {J. Maldacena, {The Large N Limit of Superconformal
Field Theories and Supergravity,} Adv.Theor.Math.Phys. {2} (1998) 231,
{ hep-th/9711200}.}

\lr\efi { G.V. Efimov, A.C. Kalloniatis
and  S. N. Nedelko, 
Confining Properties of the Homogeneous Self-Dual Field and
the Effective Potential in SU(2) Yang-Mills Theory, 
 Phys. Rev. D59 (1999) 014026,
hep-th/9806165.
}

\lr\bala {
 V. Balasubramanian, P.  Kraus, A. Lawrence and 
S.P. Trivedi, Holographic probes of anti de Sitter space times,  
 hep-th/9808017. }

\lr\adsi{
 T. Banks and M.B. Green, 
Nonperturbative effects in $AdS_5 \times   S^5$  string theory and $d = 4$ SUSY
Yang-Mills,
JHEP 9805  (1998) 002,
 hep-th/9804170;
 C.-S. Chu, P.-M. Ho, Y.-Y. Wu, 
D-Instanton in AdS$_5$ and Instanton in SYM$_4$,  Nucl. Phys. B541 (1999) 179, 
hep-th/9806103; 
 I.I. Kogan  and  G. Luzon,  D-Instantons on the boundary
 Nucl. Phys. B539 (1999) 121, 
hep-th/9806197; M. Bianchi, M. Green, S. Kovacs and G. Rossi, Instantons in supersymmetric Yang-Mills and D instantons in IIB superstring theory,
  JHEP 9808 (1998) 013, hep-th/9807033.  }

\lr\son{Y. Kinar, E. Schreiber and  J. Sonnenschein, 
$Q \bar{Q}$ Potential from Strings in Curved Spacetime - Classical Results, 
hep-th/9811192.}

\lr \gkp{S.S. Gubser, I.R. Klebanov and A.M. Polyakov, {Gauge Theory Correlators
from Noncritical String Theory,} Phys. Lett. { B428} (1998) 105, 
{ hep-th/9802109}.
  }

\lr \poly{A.M. Polyakov,  String theory and quark confinement,
{ Nucl. Phys. B (Proc. Suppl.)} { 68} (1998) 1, {{
hep-th/9711002}};  
The Wall of the Cave,
{hep-th/9809057.}}

\lr\wite{E. Witten, { Anti-de Sitter Space and Holography,} Adv.Theor.Math.Phys.
{\bf 2} (1998) 253, {hep-th/9802150};
L. Susskind  and E. Witten, {The Holographic Bound in Anti-de 
Sitter Space,} { hep-th/9805114.}}

\lr \schm{C. Schmidhuber, D-brane actions, 
Nucl. Phys. B467  (1996) 146, hep-th/9601003;
 S.P. de Alwis  and  K. Sato,
D-strings and F-strings from string loops,
 Phys. Rev. D53  (1996) 7187, hep-th/9601167.}
\lr \tte{
 A.A. Tseytlin, Selfduality of Born-Infeld action 
and Dirichlet three-brane of type IIB superstring theory, 
 Nucl. Phys. B469  (1996) 51,  hep-th/9602064.}

\lr\nahm{P.J. Braam and  P. van Baal,
 Commun.Math.Phys. 122(1989) 267;
P.  van Baal, Nahm gauge fields for the torus, hep-th/9811112. }
\lr\dougm{M.R. Douglas and G. Moore,
D-branes, quivers, and ALE instantons, 
 hep-th/9603167.}
\lr\verl{
F. Hacquebord  and  H. Verlinde, Duality symmetry of $\N=4$ Yang-Mills theory on $T^3$,
Nucl. Phys. B508 (1997) 609,  hep-th/9707179.}

\lr \othes{
J.-S. Park, Monads and D-instantons, 
Nucl. Phys. B493 (1997) 198, 
 hep-th/9612096; Z. Guralnik and S. Ramgoolam, From 0-branes to torons, 
Nucl. Phys. B521 (1998) 129,  hep-th/9708089;
 K. Hori, D-branes, T-duality and  index theory, hep-th/9902102.}

\lr\dougl{M.R. Douglas,   Branes within branes, hep-th/9512077.}

\lr\www{E. Witten, Small instantons in string theory,
\np  B460 (1996) 541, hep-th/9511030.}

\lr \chepo{  I. Chepelev and  A.A. Tseytlin, 
     Long-distance interactions of D-brane bound states and longitudinal 5-brane in M(atrix) theory, 
  Phys. Rev. D56 (1997) 3672, hep-th/9704127.
 }
\lr\hull{C.M. Hull,
Timelike T duality, de Sitter space, large N gauge theories and topological field theory,
JHEP 9807 (1998) 021,  hep-th/9806146.}

 \lr \klep { S.S.  Gubser, I.R.  Klebanov
  and A.  Peet,
Entropy and temperature of black 3-branes,
Phys. Rev.  D54 (1996) 3915, 
 hep-th/9602135; 
S.S. Gubser, I.R.  Klebanov and  A.A. Tseytlin, 
Coupling constant dependence in the thermodynamics of $\N=4$
 supersymmetric Yang-Mills theory, 
Nucl. Phys. B534 (1998) 202, 
hep-th/9805156. 
}

\lr\witten{
E. Witten, Anti de Sitter space, thermal phase transition, 
and confinement in gauge theories, Adv. Theor. Math. Phys.  2 (1998) 505, hep-th/9803131.}

\lr\suss{ L. Susskind,  Holography in the Flat Space Limit, 
hep-th/9901079.
}

\lr \kkk{ I.R. Klebanov,  From Threebranes to Large N Gauge Theories, hep-th/9901018;\ \ \ \ \ \ \ \ \ \ \ 
 M.R. Douglas  and  S. Randjbar-Daemi,
Two Lectures on the AdS/CFT Correspondence, hep-th/9902022.
}
     
\lr\izh{A. Brandhuber, N. Itzhaki, J. Sonnenschein  and  S. Yankielowicz, 
Wilson loops in the large N limit at finite temperature,  
Phys. Lett. B434 (1998) 36,  
hep-th/9803137.} 

\lr\oog{ C. Csaki, H. Ooguri, Y. Oz   and J.  Terning,
Glueball mass spectrum from supergravity. 
JHEP 9901  (1999)  017,
 hep-th/9806021; C. Csaki, J.  Russo, K.  Sfetsos  nad J. 
 Terning, 
Supergravity models for (3+1)-dimensional QCD, 
 hep-th/9902067. 
  }

\lr\isei{ K. Intriligator and  N. Seiberg, 
Lectures on supersymmetric gauge theories and electric - magnetic duality,  
Nucl.Phys.Proc.Suppl. 45B (1996) 1,
hep-th/9509066.}

\lr \haoz{A. Hashimoto  and  Y. Oz, 
Aspects of QCD dynamics from string theory, 
 hep-th/9809106.}

\lr \swa{J.H. Schwarz, An SL(2,Z) multiplet of type IIB superstrings,  Phys. Lett. B360 (1995) 13,  hep-th/9508143.  
  } 

\lr \sei {N. Seiberg,   Supersymmetry and  nonperturbative beta functions, 
Phys. Lett. 206B  (1988)  75.  } 

\lr\bbb{ E. Bergshoeff and  K. Behrndt,
D-Instantons and asymptotic geometries, 
Class. Quant. Grav. 15 (1998) 1801, 
hep-th/9803090; 
H. Ooguri and K. Skenderis,
 On The Field Theory Limit Of D-Instantons, hep-th/9810128,
JHEP 9811 (1998) 013.  }

\lr\polch{J. Polchinski, TASI Lectures on D-Branes,
{ hep-th/9611050.}}

\lr\polb{J. Polchinski, String Theory (Cambridge Univ. Press, 1998).}

\lr \hof{G. 't Hooft and M. Veltman, One-loop divergencies
in the theory of gravitation, Ann. Inst. Henri Poincare,  20 (1974) 69.}

\lr\mono{M. Li, 't Hooft vortices on D-branes, JHEP 9807 (1998) 003, 
hep-th/9803252; 't Hooft vortices and phases of large N gauge theory, 
JHEP 9808 (1998) 014, 
hep-th/9804175.}

\lr \odin{S. Nojiri and S.D. Odintsov, Two-boundaries AdS/CFT correspondence in dilatonic
gravity, hep-th/9812017.}

\baselineskip8pt
\Title{
\vbox
{\baselineskip 6pt{\hbox{   }}{\hbox
{Imperial/TP/98-99/44}}{\hbox{hep-th/9903091}} {\hbox{
  }}} }
{\vbox{\centerline {
D3-brane -- D-instanton    configuration  }  
\vskip4pt
 \centerline { and  $\N=4 $
super YM  theory } 
\vskip4pt
 \centerline { in  constant self-dual background}
}}
\vskip -32 true pt
\bigskip
\centerline{ Hong Liu\footnote{$^\sharp$}{\baselineskip8pt
e-mail address: hong.liu@ic.ac.uk} and A.A. Tseytlin\footnote{$^{\star}$}{\baselineskip8pt
e-mail address: tseytlin@ic.ac.uk}\footnote{$^{\dagger}$}
{\baselineskip8pt Also at  Lebedev  Physics
Institute, Moscow.}}

\smallskip
 \centerline {\it  Theoretical Physics Group, Blackett Laboratory,}
\smallskip
\centerline {\it  Imperial College,  London SW7 2BZ, U.K. }

\bigskip
\centerline {\bf Abstract}
\medskip
\medskip
\baselineskip10pt
\noindent
We consider $SO(4) \times  SO(6)$ invariant  type IIB string solution
describing  D3-branes superposed with D-instantons
homogeneously distributed over D3-brane  world-volume. 
In the near D3-brane horizon limit this background 
interpolates between $AdS_5 \times S^5$ space  in UV  and  flat space 
(with  non-constant dilaton and RR scalar) in IR. 
Generalizing the AdS/CFT conjecture we suggest that 
type IIB  string in this geometry is dual to $\N=4$ SYM  theory
in a state with a constant self-dual gauge field background. 
The semiclassical string representation  for the Wilson factor
implies confinement with effective string tension depending on
constant D-instanton density parameter. This provides a simple
example of type IIB string -- gauge theory duality 
with  clear D-brane and gauge theory interpretation.

\bigskip
\vskip 30 true pt


\Date {March 1999 }
\noblackbox
\baselineskip 14pt plus 2pt minus 2pt

\newsec{Introduction}

Recent  discovery   of  a  remarkable
  connection between  gravity  and gauge theory
  descriptions of   D-branes   
gives a hope  of eventual string-theoretic   description 
of properties of strongly coupled gauge theories
such as confinement \poly\ (see \kkk\ for reviews). 
The  best understood    example of this duality   
\refs{\mald,\gkp,\wite}  based on D3-branes
 relates $\N=4$ super YM theory and
type IIB string in $AdS_5 \times S^5$ space. 
To clarify the meaning of  the duality  in the context of type IIB string  theory 
 and to 
extend it to  {\it non}-conformal (and less supersymmetric)
 cases 
one may  study   various  modifications of the 
pure D3-brane  background, 
in particular:

\noindent
(i) add extra energy to the branes, i.e.   consider near-extremal D3-branes;
the corresponding gauge theory is then in a thermal  state
(see, e.g.,  \refs{\klep,\witten});
(ii)  add D7-branes  breaking supersymmetry on the gauge theory side 
to $\N=2$  \refs{\nnn,\koc};
(iii) break supersymmetry completely by adding dilaton charge
as in the solution of \refs{\kes,\gub}  (which is a 
special case of the solutions in \min\ with tachyon field 
turned off),\foot{Similar dilatonic solutions were discussed also in \odin.} 
    or, more generally, 
 exciting other scalar fields in  $S^5$-compactified 
5d theory, see  \mik\ and references there.

Here we  propose  another example  which is similar 
to the one in \refs{\kes,\gub} and \refs{\min} in that the background preserves 
$SO(4) \times SO(6)$ symmetry  but   is  simpler
having a clear interpretation  on  both  supergravity {\it and}
  gauge theory sides.
We consider  a homogeneous distribution
of D-instantons on  a large number of coincident D3-branes.
This corresponds  to  modifying the D3-brane  background by adding 
 a  RR scalar charge together 
 with (equal)  dilaton charge
in a way that preserves 1/2 of supersymmetry. 
The  resulting supergravity   solution 
is a limit of  
 superposition  \tsee\ of D3-brane  \hord\  and D(-1)-brane 
 \refs{ \gik,\bbb}  backgrounds. 
The D3+D(-1)  configuration   is T-dual to D4+D0 or D5+D1   marginal
bound state   backgrounds 
  and preserves  1/4  (1/2  near the horizon of  D3-brane)  of  supersymmetry.

In the present case  the D-instanton 
charge is homogeneously  smeared over the D3-brane   world-volume
instead of being localized  as in \adsi.\foot{A limiting 
case when  D-instanton is put at  the center 
of the  Euclidean $AdS_5$ space but is still localized   in the boundary coordinates and thus 
corresponds to a very large (i.e., approximately homogeneous)
instanton gauge  field in  the boundary  SYM theory
 was recently discussed in \suss. 
}
The corresponding gauge theory 
 is not simply  the vacuum $\N=4$ SYM theory  `perturbed' by  a single instanton
(with  perturbations 
of the supergravity D3-brane and gauge field backgrounds 
related as in  \refs{\adsi,\bala})
  but rather is  a  different     $\N=2$ supersymmetric 
state  described by a homogeneous self-dual  gauge field 
background. The latter  is 
assumed to be  averaged in some  way to preserve the Euclidean $SO(4)$
and translational  invariance. 
Its order parameter (instanton density)  is kept  fixed in the large $N$ limit.

Following \mald\  we   conjecture 
that the type IIB string theory in  near-horizon 
D3+D(-1)  geometry  is dual to  $\N=4$ SYM theory  in  
such constant self-dual gauge field background. 
This  conjecture is  consistent  
with  analogous correspondence 
between the supergravity and  gauge theory descriptions 
of D3-branes in flat space   with  additional gauge field distributions on them.
Adding a self-dual gauge field on 
a D3-brane in flat space  is equivalent  to adding D-instanton charge \refs{\www,\dougl}.
The remaining supersymmetry ensures  that the  leading ${1 \ov x^4} F^4$ 
long- and short- distance interaction 
potentials between branes (computed in classical supergravity and in one-loop SYM
theory)
agree \chept\  like  they do in the D0-brane case \dkps. This  implies 
that the (smeared)  D3+D(-1) supergravity  configuration 
corresponds   to   a D3-brane  with a (homogeneous) 
self-dual   background gauge   field.

Though   this example is somewhat  unphysical 
 (being  based on a Euclidean background   which usually  
has a well-defined meaning only
 as a virtual  configuration in the path integral)
 our main motivation is that it provides a clear
setting  for testing  certain aspects
 of string theory -- gauge theory  duality.
As in other  similar  cases   \refs{\min,\kes,\gub,\mik}
here the background  interpolates  between  a regular  UV \ 
$AdS_5 \times S^5$  space
(conformal gauge  theory) 
  and 
a singular IR  background.  
A remarkable simplification  which happens in  D3+D(-1) case  
(in addition to a transparent
analytic form of the supergravity solution) 
is that the  IR limit of the string-frame 
metric is  actually  flat,\foot{The IR limit 
of the Einstein-frame metric is  again  $AdS_5 \times S^5$.}
 with the 
singularity of the background residing only  in the dilaton  field.
Assuming  the  string-theory representation for the Wilson factor 
 \malr\ that immediately leads
(in the semiclassical approximation) 
 to the area law behaviour.

This prediction of the  `string in curved space -- gauge theory'  duality is 
consistent with expectations that  gauge theory in a constant 
self-dual background \refs{\mink,\leit,\oth,\efi} 
should be (partially)  confining. The confinement should be  simply  a   consequence
 of the presence of a  non-trivial    background  field
(implying, e.g., that  propagators
do not have poles in the complex energy plane \efi).
We find that the effective string tension 
following from the  curved space picture 
  depends indeed only
on the background  gauge field  order parameter  and not 
on the gauge coupling.

We shall start in section 2  by presenting   the 
D3+D(-1) supergravity 
background. 
The corresponding 
gauge theory state  described  by a 
self-dual background   will be considered  in section 3. 
The  relation between  with the  supergravity solution
and gauge theory 
state will be  illustrated   in section 4 by 
SYM   effective action interpretation of
 the D3-brane probe action in  the D3+D(-1)  background.
In section 5  we shall use the dual  string theory 
description 
to compute the Wilson loop factor using the semiclassical fundamental string action 
and  demonstrate  that it exhibits  area law behaviour.
In section 6 we shall present  analogous 
discussion of the `t Hooft loop factor  based on the 
D-string action and show that magnetic monopoles are 
screened at large separations. 
Section 7 will contain some concluding remarks.
In Appendix we shall 
consider perturbations of the background fields
 near the D3-brane + D-instanton supergravity background and 
demonstrate that scalar field perturbations  do not, in general,  
decouple from the metric one. 
While one linear  combination of the dilaton and RR scalar
perturbations  
still satisfies the massless scalar equation in $AdS_5 \times S^5$
(so there are still gapless excitations in this background),  
the second combination mixes with graviton and should 
have complicated dynamics, 
 reflecting  the presence of the 
self-dual  gauge field  background on the  gauge theory side.

\newsec{Supergravity  picture: 
D3+D(-1) background }
A simple  example of  a `deformation' of the D3-brane solution 
of type IIB supergravity  by   non-trivial  scalar 
field backgrounds is obtained 
by switching on a RR scalar charge and balancing 
it  by a dilaton charge to preserve 1/2 of supersymmetry. 
The resulting exact solution  representing 
a  marginal `bound state' of D3-brane  and 
a (smeared) D-instanton   
is 
a non-trivial  example of a  
  type IIB   brane intersection configuration  
parametrised by two harmonic functions. It has the following   
  string-frame metric  \tsee\
\eqn\nst{  ds^2_{10}= (H_{-1})^{1/2}  
( H_3^{-1/2}  dx_m dx_m + H_3^{1/2} dx_s dx_s )\  ,  } 
where $m=1,...,4$,\ $s=5,...,10$, and  $H_{-1} $ and $H_3$ are the 
 harmonic functions depending only on  $x_s$.
In addition to the self-dual 
 5-form background ($C_{1234}= H^{-1}_3$) representing the
  D3-brane charge we have the  non-trivial 
dilaton $\P= \p_0 + \p$  and RR scalar backgrounds\foot{We  absorb
 the asymptotic value of the dilaton  \ 
($e^{\p_0}=g_s$) into the overall coefficient ${1 \ov 2\kappa^2} = { 1 \ov (2\pi)^7 g^2_s \a'^4}$ of the 
action (rescaling also  the RR fields by $g_s$-factor).
Here  $\C=i C$,  where $C$ is the original 
 RR scalar. In the 
  type IIB theory with the 
 Euclidean time $x_4= it$ the  self-dual RR 5-form  is imaginary, 
so this  solution  is   complex in both
 Euclidean and Minkowski versions. We shall view D-instanton 
or D3+D(-1) backgrounds    as formal complex solutions of  type IIB theory.}
\eqn\diim{
 e^{\p} =  H_{-1} \ , \ \ \ \ \ \ \ \   \C=  (H_{-1})\inv   -1   
\  . }
This solution is  T-dual to $D4\pa D0$ or $D5 \pa D1$ configurations.
The total  stress tensor of the two scalar  fields 
is zero so that  the Einstein frame metric remains the same as in the D3-brane case. 

Since the D-instanton is smeared  over the 3-brane  world-volume 
 the two  harmonic functions have the same  structure ($r^2 = x_s x_s$) 
\eqn\harm{
H_3 = 1 + { Q_3 \ov r^4} \  , \ \ \ \ \   \ \ \  
H_{-1}  = 1 + { Q^{(4)}_{-1}  \ov r^4} \ , }
where the coefficients  are normalized as follows (see, e.g., \refs{\polch,\chept})\foot{ In general,  $T$-duality implies that 
for a 
Dp-brane  background which is smeared in $n$ transverse toroidal directions
$H_p = 1 + {Q^{(n)}_p\over r^{7-p-n}}$,
where $Q_p =  N_p g_s (2\pi)^{(5-p)/2} T^{(p-7)/2} (\omega_{6-p})\inv
  , \  $ $
 \omega_{k-1} = 2 \pi^{k\ov 2}/\Gamma({ k\ov  2})  ,$  $  \
Q^{(n)}_p 
  = N_p N_{p+n}\inv  Q_{p+n} (2\pi)^{n\ov 2} T^{- {n\ov 2}} V_n\inv $. 
Here   $V_n$ is the  volume of the flat internal  torus and $T= {1\ov 2 \pi \a'}$.}
\eqn\norm{
Q_3 = 4\pi N g_s \a'^2 = \l \a'^2 \ , \ \ \ \ \ \ \ \ \ \ \ \ 
Q^{(4)}_{-1} = {\ni \ov N} { (2 \pi)^4   \a'^2\ov  V_4}  Q_3
=    q \l \a'^4 \ ,  }
\eqn\fii{
q \equiv  {\ni \ov N} { (2 \pi)^4 \ov  V_4} \ , \ \ \ \ \ \ \  
  \l= 4\pi N g_s  \ , \ \ \ \ \ \ \ \  e^{\p_0} = g_s \ .   }
The integers $N$ and $K$ are the  D3-brane and D-instanton numbers.

The supergravity approximation   applies 
 when  derivatives of the fields
remain small in  units of $\a'$, i.e.  
\eqn\deri{ \l \gg 1 \ ,  \ \ \ \ \ \ \ \  r \gg  (q\l)^{1/5}\a'^{9/10}  \ ,   }
and when the effective string coupling is small, i.e. $ r \gg (q\l)^{1/4}{  \a'}$. 

Let us now consider the  following generalization 
of the decoupling limits of \refs{\igo,\mald,\itzh}
\eqn\limi{
\a'\to 0\ , \ \ \ \  \{ u = {r \ov \a'}, \  \l, \  q \}= {\rm fixed} \ . }
This limit is equivalent to taking a special 
solution without 1 in the 3-brane  harmonic function $H_3$
(and making a coordinate redefinition).
The  dilaton and  RR scalar are  expressed \diim\ 
in terms of  the function $\hi$  which  remains  finite in this limit\foot{This 
limit was considered also   by N. Itzhaki.} 
\eqn\hii{
   \hi = 1  + { q \l \ov u^4} \ ,  }
while 
 the  string-frame metric  takes the following form
\eqn\nit{  ds^2_{10}= 
\a' \l^{1/2}  (1  + { q \l \ov u^4})^{1/2} 
\bigg( \l^{-1} { u^2}   dx_m dx_m +  {du^2 \ov u^2}  + d\Omega_5^2 \bigg)\  .  } 
The corresponding Einstein-frame metric 
$ds^2_E = e^{-\p/2} ds^2_{10}$  is  still 
$AdS_5 \times S^5$ as in the D3-brane case.
 As in the
 pure D3-brane case \mald, 
the factor $\a'$  cancels out  in the  {\it classical} 
string action 
  which thus remains finite in the limit \limi.
 
Like  the $AdS_5 \times S^5$ 
Einstein frame metric, the string frame metric \nit\ 
is completely  regular: near the core  $r\to 0$ (i.e.  in the IR region $u\to 0$)
 it  becomes  flat   
\eqn\nitk{  ds^2_{10}= 
\a'  q^{1/2} 
 \bigg(   dx_m dx_m +   d z ^2   + z^2 d\Omega_5^2 \bigg)\  , \ \ \ \ \ \ \ \ 
z\equiv  { \l^{1/2}\ov  u}  \ . } 
 Note that the overall coefficient (effective string tension)
 is  just 
$q^{1/2}$, i.e.   does not depend on $\l$.
 The  
full D3+D(-1) background is, however, 
 singular because of the  
 blow-up of  the  dilaton at $u=0$.
This is the same  behaviour as in the case of the  
 pure  D-instanton  background  \gik.

The background  \nit\  thus interpolates between 
 $AdS_5 \times S^5$ at  $u=\infty$ (UV) and 
flat space with singular dilaton   near $u=0$ (IR), 
while in the absence of D-instantons ($q=0$) 
the  $AdS_5 \times S^5$  space describes both regions.
The constant instanton density parameter $q$ breaks 
conformal invariance in the IR region. 
Similar interpolating backgrounds  (with singular metrics in the IR) 
were found in \refs{\min,\kes,\gub,\mik}.
An attractive feature  of the present case is its 
simplicity (not unrelated to remaining supersymmetry)
and clear identification  of the corresponding dual gauge  theory   --
$\N=4$ SYM with  a constant self-dual  background (section 3). 
This interpretation is supported  by  the relation  between 
 the classical action of a  D3-brane probe moving  in this geometry 
and  the  SYM  one-loop effective action in the corresponding gauge field 
background
(see section 4). 

The type IIB string theory in the background \nit,\diim\
is thus conjectured to be dual to a certain $\N=2$ supersymmetric,  
and $SO(4) \times SO(6)$ and translationally invariant  state 
of $U(N)$ $\N=4$ SYM theory (which may be  equivalent to  some  $\N=2$ supersymmetric YM theory with matter which is conformal in 
the UV). 
Symmetries of the string theory should 
therefore  imply certain symmetries on the gauge theory side. 
The string theory and 
gauge theory   are parametrised  by $\l,\ N, \  q, \  V_4$ 
(or  $g_s, \ N,\ K, \  V_4$).  

  One obvious string symmetry is $T$-duality.
The original  D3+D(-1) background \nst,\diim\
with standard ($r\to \infty$) asymptotically flat region
is  covariant  under T-duality along
all four  3-brane world-volume directions: 
$T$-duality  
interchanges
$H_3$ with $H_{-1}$, i.e. $Q_3$ with $Q^{(4)}_{-1}$.\foot{More precisely, 
here under $T$-duality \ $ V_4 \leftrightarrow \td V^{(0)}_4,$  $\  g_s \leftrightarrow \td g^{(0)}_s,$  $\  N \leftrightarrow K$, where $ V_4 \td V^{(0)}_4 = (2\pi)^8 \a'^4$, 
\ $V_4/g^2_s =\td  V^{(0)}_4/\td g^{(0)2}_s$,
and  
$\td g^{(0)}_s = { (2\pi)^4 \a'^2 \ov V_4} g_s$ is the dual coupling in
 the standard asymptotically  flat region.  As a result (see \norm)\  
$\td Q_3 =   Q^{(4)}_{-1},\ \  \td  Q^{(4)}_{-1} = Q_3$.
}
The limiting background 
\nit,\diim,\hii\ (and $C_{1234}=H_3\inv$)
is parametrised by  `asymmetric' combination of harmonic functions 
($H_3' \equiv  H_3-1=  { \l\ov \a'^2 u^4}, \ \ H_{-1} =1 + { q \l \ov u^4}$)
and thus changes its form under the  $T$-duality.
 Assuming  that all four directions are 
compact with equal radii $a$,\ 
 $x_m = a \theta_m $,  \  $\theta_m \equiv \theta_m + 2\pi$\  
 (i.e. $V_4 = (2\pi a)^4$)
 and that the dual angular coordinate  is 
$\td \theta_m \equiv \td \theta_m + 2\pi$, 
 the  string metric 
$T$-dual to \nit\ is found to be 
(in terms of the string Lagrangian   
${ 1 \ov 4\pi \a'}  G_{10} (\del  x)^2 + ...\ 
 \to \ { 1 \ov 4\pi \a'} \td G_{10} (\del \td x)^2 + ... $)
\eqn\nitdu{  {d\td s}^2_{10}= 
\a' \l^{1/2}  (1  + { q \l \ov u^4})^{1/2} 
\bigg[  { d\td \theta_m d\td \theta_m \ov a^2  u^2 (1  + { q \l \ov u^4})} 
   +  {du^2 \ov u^2}  + d\Omega_5^2 \bigg]\  .   } 
The   effective string coupling  $e^{\td \P}$  
dual to $e^\P = g_s H_{-1}= g_s (1 + { q\l\ov u^4})$ 
is
\eqn\dila{
e^{\td \P} = \td g^{(0)}_s H'_{3} = { \l \ov a^4 u^4} \ . } 
The large $u$ limit of the dual  metric \nitdu\  still has the same 
$AdS_5 \times S^5$ form (with $\td x_m = a \td \theta_m$)
  when  written 
in terms of the `dual coordinate' (or `dual energy scale') 
$\td u= {  \l^{1/2} \ov  a^2} { 1 \ov  u}$.\foot{This is the same 
conclusion as in the near-horizon 
 D3-brane case 
which is mapped by $T$-duality 
into the near-horizon limit of smeared D-instanton \kall. Though the 
$AdS_5$ metric with compact $x_m$ is singular at $u=0$, string theory is supposed to resolve this singularity  and should  be dual to  SYM  theory 
on  a 4-torus. }

The  small $u$ limit of \nitdu\  is again a flat space region 
(cf.\nitk).  Using the expression for $q$  \fii\ 
 the fundamental string  Lagrangian  in the 
near $u=0$ region \nit\  can be  put into the form 
$L = { 1 \ov 4 \pi} \sqrt{ K \ov N}\ \del_b \theta_m \del^b  \theta_m + ... $
(the radius $a$ cancels out). The  $T$-dual  string Lagrangian
is then 
\eqn\duaa{
L = { 1 \ov 4 \pi} \sqrt{ N \ov K}\  \del_b \td  \theta_m \del^b
 \td  \theta_m + ... \ ,}
 i.e. the duality corresponds to interchanging 
$K \leftrightarrow N$. 
 Measured in terms of the string metric 
the dual radius is $\td a_s = { 1 \ov \sqrt q a} $ 
 (near $u=0$  the effective string tension
is $\sqrt q \ov 2\pi$, see \nitk).
The   gauge theory  volume is 
defined, however, 
 as the volume of the boundary theory 
 in  the $u\to \infty$ $AdS_5\times S^5$ space  limit and 
remains  unchanged, $\td V_4 = (2\pi a)^4$ (we consider the simplest rectangular 4-torus).
Note that  near $u=0$  and $u=\infty$ 
 the ratio
of the  effective string couplings ${e^{\td \P}\ov e^{\P}} \equiv 
 {\td g_s(u) \ov g_s(u)}$   becomes 
$
{\td g_s(0) \ov g_s(0)}  = { N \ov K}   ,  \
 \ 
   {\td g_s(\infty ) \ov g_s(\infty)}
 = { \l \ov a^4 u^4  }|_{u\to \infty}  \to 0 .$

We  are thus 
  led to the conclusion that string $T$-duality 
implies an equivalence 
relation between  $U(N)$ gauge theory state  parametrized by
$K$ (and  $g_s(u)$)  and $U(K)$ gauge theory state parametrized by
$N$  (and $\td g_s(u))$. 
 Given that $K$ may be interpreted (section 3) 
as an  instanton number in $U(N)$ gauge theory, 
this suggests a  relation  to  the   Nahm duality
which maps  a charge $K$,\ $U(N)$\  self-dual solution on a 4-torus
to a charge $N$,\ $U(K)$\  self-dual solution on a dual  4-torus \nahm.
Nahm duality was previously 
discussed in connection with $T$-duality in string theory  in 
{\it flat} background in, e.g., 
\refs{\dougm,\verl,\othes}. In the present 
context of the ``curved space 
string theory --  gauge theory'' duality 
conjecture  
  one  indeed seems  to see  why  
``YM theory (via the Nahm transformation)
 still knows about $T$-duality", 
cf. \verl. 
Here we  are getting  
not  just a relation  between two classical self-dual 
Yang-Mills  backgrounds, but  an equivalence
between two quantum  gauge theories 
(which is reminiscent of some 
 known  dualities between supersymmetric 
gauge theories \isei).

The forms of the dilaton  and RR scalar \diim,\hii\ 
 suggest the following
 expressions for the  IR  flow   of the SYM couplings 
\eqn\ruu{
 \l(u) = \l ( 1  +  { q \l\ov u^4} )  \ , \ \ \ \ \ \ \ 
\theta(u) = ( 1  +  { q \l\ov u^4} )\inv -1 \ .  
}
Similar power-like running 
of the dilaton was found also in \refs{\kes,\gub}.
In the $\N=2$  gauge  theory  related to  $D7+D3$ brane  configuration  \nnn\ 
analogous   IR running of dilaton  may be  explained by instanton  corrections 
(see \koc\ and refs. there). Here it should be induced 
by the presence of a constant self-dual field 
condensate.\foot{Instanton contributions
do not produce   IR \ 
 RG flow  
  in the $\N=4$ SYM theory in the  vacuum state \sei.}

The flatness of the $u\to 0$  limit of the metric 
  suggests that the IR  limit of gauge theory should be described by 
type IIB string theory in {\it flat} space (with additional dilaton and RR scalar and 4-tensor backgrounds  being invisible in a semiclassical approximation).
That obviously leads to the   area law behaviour of   the 
Wilson factor (see  section 5).

The major question, however, is 
 whether  the background \nit,\diim,\hii\  may  be extrapolated 
all the way into the IR region $u\to 0$. 
We  expect  that due to its remaining supersymmetry 
 this  
background, like $AdS_5 \times S^5$, 
 is  actually an exact (all order in $\a'$)   solution of 
the classical type IIB string theory.
  Then  
 one may relax the conditions \deri\ but 
there is still the condition that 
the  effective string coupling $e^\p$ should 
 remain  small, i.e.  
\eqn\bou{
u  \geq    u_*\  ,  \ \ \ \ \ \ \ \ \ \  u_* \gg   (q \l)^{1/4} = 2 \pi ({\ni \l \ov N V_4})^{1/4}  \ . }
If $q\l$  is  not constrained,  we may 
allow $u_*$ to be sufficiently 
small  to probe the IR limit  and still have weak  
string coupling  to ignore string loop corrections.
This will be our assumption below.
 $u_*\sim (V_4)^{-1/4}$ may be interpreted as an IR  energy cutoff 
on the gauge theory side.

\newsec{Gauge theory  picture}
The gauge field theory counterpart of the 
supergravity solution discussed  above 
is  the  $\N=4$  $SU(N)$ SYM theory  in a  certain (non-vacuum) state 
which may be  described 
as a stochastic averaging of   a constant homogeneous self-dual  background  
(see, e.g., \refs{\mink,\leit,\oth,\efi}).\foot{Related homogeneous distributions
of instantons were considered also in \kli.}
 Namely, we shall 
demand  that in such state\foot{Here $\tr$ is in the fundamental representation, $\tr (T_aT_b) = \ha \d_{ab}$.  
The classical action  is  $\ \ \ \ \ \ \ \  \ \ \ \ \ \ \ \ $
  \ \ \ \ \ \ \ \ $  \, \, \, \, \, \qquad \ \ \  \ \ \ \ \ \ \ \ \
 S= { 1 \ov 2 \gym^2 } \int d^4 x\  
\tr ( F_{mn} F_{mn})   - { i \theta \ov 16 \pi^2} \int d^4 x \ 
\tr ( F_{mn} F^*_{mn}) \ .$  
} 
\eqn\den{
< F_{mn}> =0\ , \ \ \ \
 < \tr (F_{mn}F_{mn} - F_{mn}  F_{mn}^*) > =0 \ ,
 \ \ \ \ \ \ \ \ 
 < \tr ( F_{mn} F_{mn})  >  = { 16 \pi^2  \ov V_4}   \ni   \ , }
where   a finite volume $V_4$  will be   used as an IR cutoff. 
We shall assume that this $N=2$ supersymmetric state  may  obtained by 
some  averaging over different 4d directions 
  so that  $K$ is the only order parameter, i.e. that  the 
 Euclidean  group -- $SO(4)+$translations --  is unbroken 
 (while P- and T- invariances are   obviously broken).
Indeed, the supergravity solution has the same 4d Euclidean group  
symmetry 
and depends only on one extra constant $q$ which is related to
 $K$.

For  the   purpose of comparing with supergravity
 such  state  may be represented simply  by  (a stochastic averaging of)  
a constant  self-dual abelian background  such that 
\eqn\deni{
 < \F_{mn} \F_{mn} >   =< \F_{mn} \F^*_{mn} > = 
 {1 \ov N} < \tr ( F_{mn} F_{mn}) >  = { 16 \pi^2  \ov V_4} 
   { \ni \ov N }  \ . }
We shall consider the large 
 $N$ limit defined by 
\eqn\lan{
N, \ni, \gym^{-1} \to \infty, \ \ \ \  \ \ 
 \l\equiv  N \gym^2 = {\rm fixed} \ , }
\eqn\pop{ \ \ \ \
q \equiv { (2\pi)^4  \ov V_4} 
   { \ni \ov N }  = {\rm fixed} \ ,  
\ \ \ \ \ \ \ \ \  q = \pi^2 < \F_{mn} \F_{mn} >  = \const\ . }
Again, 
 $q$ will be the only  order parameter 
 `visible' on the supergravity side, i.e. 
the gauge theory state is  defined by averaging  over all fields
satisfying $\F_{mn}= \F^*_{mn}$ and \deni.\foot{Similar   averaged gauge field configurations
were discussed in the  context of comparing with supergravity
solutions in, e.g.,  \refs{\mmm,\chee}.
The idea is to `engineer' a gauge field background
that has certain  global characteristics (energy, momentum, instanton 
number, etc.) which are the only parameters  present
 in the supergravity solution. } 

Introducing a  background field does not 
 change  UV finiteness  property of the $\N=4$ SYM 
theory, 
 but  may, of course, 
modify IR behaviour. 
Indeed, the propagators of massless fermions
and `off-diagonal' gluons in the self-dual  background 
exhibit  confinement-type (no-pole)  behaviour \efi.
The confinement on the SYM side is  
caused   by the background field 
 \refs{\oth,\efi}.
The dual string/gravity description 
will lead to  an  area law for the Wilson factor with 
effective  string 
tension  proportional simply  to  $\sqrt q$ (section 5).
  The supergravity description 
also predicts  \ruu\ 
that 
 the  gauge coupling constants ($\gym, \theta$)  should 
have power-like running in the IR.

\newsec{D3 brane probe in D3+D(-1) background  and SYM effective action} 
To  support the gauge theory interpretation
of the  above supergravity  background
let us consider the action of a D3-brane probe moving 
in the D3+D(-1) background  and demonstrate  its 
relation to the quantum  effective  action of $\N=4$   SYM 
theory. The latter will be computed  in a constant gauge field  background 
containing self-dual component  on the stack of $N$ D3-branes 
representing a source    and a constant  abelian field
on  a separated  D3-brane probe  (see, e.g.,  \refs{\dougli, \chept,\mmm,\chee,\mald,\dougtay}). 
We shall    choose 
the  static gauge on the Euclidean D3-brane  world volume action. 
 Then the action for a D3-brane probe moving
in the background \nit,\diim,\hii\ and $C_{1234} \sim u^4$ takes the form\foot{ 
  We use that the D3-brane  tension is 
$T_3= { 1 \ov 2 \pi  g_s (2\pi \a')^2}$ and absorb the factor 
$2\pi$ into $F_{mn}$.} 
$$
I_3 =  { N  \ov 2 \pi^2 \l}  \int d^4 x
\bigg[
   u^4   \sqrt{ \det\bigg( \d_{mn} 
 + {  \l\ov  u^{4}}  \del_m u^s \del_n u^s + 
  {  \l^{1/2} \ov   (u^4 + \l q)^{1/2}}  F_{mn} \bigg) }
  $$
 \eqn\probe{  - \ u^4 \ 
 +    \     {  \l q \ov 4(u^4 + \l q) }   F_{mn} F_{mn}^* \bigg] \ , 
}
where we have  used \diim,\hii\ in the RR scalar coupling term 
$-i C F_{mn} F_{mn}^* = -\C F_{mn} F_{mn}^*$.
Expanding in powers of derivatives we get\foot{The leading interaction term here  vanishes  
in the case of the self-dual $F_{mn}$  which corresponds 
 to adding D-instanton charge
to the D3-brane probe. This is the expected cancellation
of the  static  potential 
between the  same-type  marginal bound states of branes like
$D3+D(-1)$  and $D3+D(-1)$  or $D4+D0$ and $D4+D0$
\refs{ \chepo,\chept}.}
\eqn\expa{
I_3 =  { N  \ov 2 \pi^2 }  \int d^4 x
\bigg[ \ha \del_m u^s \del_n u^s  + \fourth    F_{mn} F_{mn} 
   - {  q \ov 4(u^4 + \l q) } 
 ( F_{mn} F_{mn}   -   F_{mn} F_{mn}^*)  + ... 
 \bigg] \ . 
}
This action can indeed be interpreted  as a 
result of  integrating out  open string modes  connecting 
the  probe and the source  separated by distance $u$ in flat space 
with $F_{mn}$ background on the probe and 
a self-dual  background parametrized by $q$ on the source.
Let us  assume that $N$ coinciding  D3-branes  carry 
 a constant $U(1)$ self-dual background $\F_{mn}=\F_{mn}^*$,
so that the total $U(N+1)$ 
background  
is    block-diagonal, 
$\FF_{mn} = \diag (\F_{mn} I_{N\times N}, F_{mn}  I_{1\times 1})$.
The leading  $F^4$ one-loop correction in the 
$\N=4$ SYM effective action   has the following structure \ft
\eqn\ymm{
 { 1 \ov  u^4} {\rm STr }  \bigg[  ( \FF_{mn} \FF_{mn})^2   -  ( \FF_{mn} \FF_{mn}^*)^2 \bigg]
\ , } 
where $\FF$ is the total background field, $u$ is the effective mass scale 
set by the separation between branes
and $ {\rm STr }$ is the symmetrized trace in the adjoint representation.
 For  a commuting 
  block-diagonal  background 
$\FF= \diag (\F, F)$ 
we  may  simply to replace $\FF_{mn}$ by $F_{mn} -\F_{mn}$ under the adjoint trace
 adding the overall factor (number of non-zero eigenvalues)  $2 N $.
Then the  $\F^4$ term  in \ymm\ vanishes  because of the  self-duality of $\F$ and 
 the $ O(\F^2 F^2) $  cross-term   becomes  proportional to\foot{Note that 
the  $O(\F F^3)$ term 
$ { N \ov u^4}  \F_{mn} F_{mn}   ( F_{mn} F_{mn}   -   F_{mn} F_{mn}^*) $
absent in the supergravity expression 
 vanishes  after averaging over 
the directions of the self-dual field $\F$, 
implying $< \F_{mn}>=0$.
As was already mentioned above,  this averaging is necessary 
  in order for the  brane source gauge theory 
state to represent the  $SO(4)$ 
invariant  supergravity background. The $O(F^4)$ term in \ymm\ is also reproduced by
the $F^4$ term in the expansion of \probe.}
\eqn\cros{
{ N \ov u^4}\  \F_{mn} \F_{mn} \  ( F_{mn} F_{mn}   -   F_{mn} F_{mn}^*) 
 \ . } 
 Assuming  that  $\F_{mn} \F_{mn} = q/\pi^2$  as in \pop\ 
and expanding \expa\ in powers of $q$ 
we find the  (precise numerical) agreement 
between  the supergravity \expa\ and the  SYM  \cros\ expressions
(see \chept\ for details
of  similar calculations).\foot{From the 
`D-branes in flat space' perspective the reason why 
the supergravity (long-distance) and SYM (short-distance)  
 interaction potentials    agree \dkps\ 
is in  the non-renormalization theorem for $F^4$ terms \dine.}
A non-trivial 
feature of this example 
is that while  
$q$ and $F_{mn}$  have very different  interpretation on the supergravity side
(one is a parameter of the background and another is a `coordinate' 
of the  probe)
they have the same  background gauge field interpretation
on the gauge theory side.

\newsec{String representation for the  Wilson  loop}
Let us now  consider  a  type IIB string propagating 
in the D3+D(-1)  supergravity background 
and  compute, following \refs{\malr}, 
the potential between  far separated 
`quark' and `anti-quark'  (W-bosons) by evaluating
the string partition function representation for the Wilson factor 
 in the semiclassical  approximation.
This amounts to computing the value of the  
bosonic Nambu string action on the classical string configuration.
Assuming that the string penetrates the  IR ($u=0$) region 
the flatness of the string metric   in this region \nitk\
implies 
that the Nambu action will be proportional 
to the area. This  implies the 
confinement behaviour 
similar to what is found  in the non-extremal (finite temperature) 
case \izh\
 and  in the  case of the  dilaton-charge deformed 
 3-brane  solution  \gub.

We shall take the fundamental 
 string action in the form $ { 1 \ov 2 \pi \a'} \int  d^2 \s
\sqrt {\det  G_{ab} } $ 
and fix the static gauge  $x^0 =  \tau, \ \  x^1  =  \sigma$,
where 
the world-sheet coordinates $\tau$ and $\s$ 
 run  from 0 to $\T $ and $-L/2$ to $L/2$.
As in  \malr,  we shall    consider  a static 
 string solution  which  has only the radial coordinate  
 $u$ changing with $\s$.
The  stationary point is  found 
 with the boundary condition  that  $u$ 
 runs to  infinity at $-L/2$  and  $L/2$
and 
takes  the 
minimal value  in between.
Since  the  dilaton and RR field couplings  can be ignored 
in the semiclassical approximation, the string  action is 
determined  simply by the  metric \nit\ 
\eqn\stre{
I=  { \T \ov 2\pi     }  \int d \s   \sqrt { 
  H_{-1} (u) 
   \ (  u'^2  +  \l^{-1} u^4) } \ ,  \ \ \ \ \ \ \ \  \ \ 
 H_{-1}  = 1  + { q \l \ov u^4}\ , 
}
or, equivalently, 
\eqn\stret{
I =   { \T   \ov 2\pi     }  \int d \s\   n(z)   \sqrt { 
     z'^2 +  1 } \  , 
}
\eqn\zz{ n(z) = \sqrt{ q  + {\l \ov   z^4} } \ , \ \ \ \ \  \ \ \ \ \ \   z = {\l^{1/2}  \ov u} 
\ . } 
Since $f^2(u) \equiv H_{-1}\l\inv u^4  = \l\inv u^4 + q  $ 
has a minimum at $u=0$  it follows 
 from the general analysis of \son\ that 
the action \stre\  leads  to
 confinement with the string tension  proportional to 
$f(0) = \sqrt q$. The same conclusion 
follows from the light-ray analogy used in \gub\  to argue
that for a  large  separation $L$  the potential  is dominated by 
the minimum of the `refraction index' $n(z)$, i.e. by 
the limit $z\gg (\l/q)^{1/4}$.\foot{Note that in contrast to the  case  discussed in 
 \gub\  here the  effective tension does not depend on $\a'$.}

Indeed, the large $L$ limit is dominated  by the IR region  $u\to 0$ 
where the string-frame metric \nit\ becomes flat \nitk,  so that 
the string action \stret\  takes the flat-space form
\eqn\street{
I\  \to \    { \T  \sqrt q \ov 2\pi     }  \int d \s \    \sqrt { 
     z'^2 +  1 } \  ,   
}
and thus automatically 
 leads to the area-law  behaviour of the Wilson factor, 
\eqn\ret{
W(C) \  \to  \  e^{- T_{\rm eff} { \T  L}  } \ , \ \ \ 
 \ \ \ \ \ \ \ \ \ \
 T_{\rm eff} =   {  \sqrt q\ov 2\pi}  \ .   }
An 
 interesting feature is that the effective string 
 tension $T_{\rm eff}$ does not depend on the  `t Hooft coupling
 $\l$, but depends  only on the  background  order    parameter  $q$
 (instanton density).
This is  consistent   with  
 the gauge theory picture (section 3)  
since   confinement in a self-dual gauge field 
 background   should be  caused simply  by the  gauge field condensate.

In  the above discussion  
we  have    ignored the fact that  near the true minimum of $n(z)$, i.e.
 $z=\infty, \ u=0$, 
the string coupling  becomes  strong  and thus string loop corrections
(higher topologies) may  not be, in principle, ignored.\foot{This problem did not appear in the dilatonic charge solution \refs{\kes,\gub} 
   where the supergravity approximation was still valid 
near the minimum \gub.}  
For large $L$ 
 the string probes  the far IR region where  the string coupling is no
longer  small.  A  way  to  avoid  this problem  may be to 
impose the IR cutoff \bou\  $u \geq u_*$, i.e.  to  restrict $z$ to  values
$z \leq   z_* \ll (\l/q)^{1/4}$. 
Then the minimal value  of $n(z)$ will be  at $z=z_*$ 
and the potential  will contain additional dependence
on $L/z_*$. 
 One may qualitatively approximate the resulting 
 effective string tension  as
$
T_{\rm eff} =  { 1  \ov 2\pi}
\sqrt{ q  + {u^4_*  \ov \l  } }$ which reduces to \ret\ in the limit 
$u_*\to 0$. 
 
\newsec{D-string action, S-duality  and `t Hooft loop }

Additional insight into the structure of the  duality 
 is obtained by  probing  D3+D(-1) background  with a D-string.
 Following \refs{\mono,\gross} the exponential of the 
D-string action  determines (in the semiclassical approximation)
the corresponding `t Hooft loop factor  and thus 
the  potential between magnetic monopoles on the  gauge theory side.

In general, a  D-string  action in a  background 
  with  a non-trivial metric, dilaton and RR scalar 
has the form\foot{Here $T_1 = { T \ov g_s } = { 1 \ov 2\pi \a' g_s}$
and we absorb the $2\pi \a'$ factor into the Euclidean world sheet 
 gauge field $F_{ab}$.}
\eqn\ddew{ I_{1} = T_1 \int d^2 \s \ \big[ \ e^{-\p} \sqrt {\det( 
 G_{ab}  + F_{ab}) } + 
{ \textstyle {i\ov 2}} \ep^{ab} C F_{ab}\ \big]   \ . }
Solving for the 
2d  gauge field $F_{ab}$, i.e. `integrating it out' in the 
semiclassical approximation (as, e.g.,  in \refs{ \schm,\tte}) 
gives   
\eqn\ddew{ I'_{1} = T_1 \int d^2 \s \ 
\sqrt{  e^{-2\p}  + C^2} 
\sqrt {\det G_{ab} }  \ .  }
The difference as compared to the fundamental string Lagrangian is thus
in the  `tension factor'  which in the case of the D-instanton background  \diim,\hii\  becomes ($\C=i C$)
\eqn\dife{
{ 1 \ov g_s} 
\sqrt{  e^{-2\p}  - \C^2}  = 
{ 1 \ov g_s} 
\sqrt{  2 e^{-\p}  - 1} = { 1 \ov g_s} 
\sqrt{ 1 - { q \l\ov u^4} \ov 1 + {q \l\ov u^4} } \ . }
The D-string counterpart of the  fundamental string action  \stre\ is thus 
is 
\eqn\dre{
I=  { \T \ov 2\pi g_s    }  \int d \s   \sqrt { 
  ( 1 - { q \l\ov u^4} )  \ (  u'^2  +  \l^{-1} u^4) } \ . }
Remarkably, 
this action is exactly the same (up to the tension ratio factor $1/g_s$)
as the fundamental string 
 action \stre\ but   with $q \to -q$ !  
Since the F-strings and D-strings 
are interchanged by S-duality \refs{\swa,\schm,\tte}, 
 this suggests that  $q \to -q$ 
is simply a consequence of  S-duality transformation 
applied to  our background --
D-string  `sees' S-dual geometry \nit,\hii,\diim\  with $q \to -q$. 

Indeed, under the basic  S-duality transformation  
$\tau \to - { 1/ \tau} ,  \  \tau = C + i e^{\p},$  which is 
 a symmetry of the scalar $SL(2,R)/SO(2)$ sigma model 
 action  $( \del \p)^2 + e^{2\p}( \del C)^2$ one has 
\eqn\dfd {  e^{  \p'} =    (e^{-2\p} +  C^2 )\ e^{\p} \ , \ \ 
\ \ \  \ \  C' = - { (e^{-2\p} +   C^2)\inv  C  } \ ,  \ \ \ \ \  \ 
  g'_{\m\n E} = g_{\m\n E} \ .   }
This target space transformation  combined with world-volume 
duality  is a symmetry of the  D3-brane action \tte\ and 
relates D-string and F-string actions. 
In the case of the 
 D-instanton or D3+D(-1)  backgrounds \diim,\hii\    one finds from \dfd \foot{We stress that this  
S-duality is the standard  transformation of the 
Minkowski type IIB supergravity with $C$ formally replaced  by 
$i\C$ in the case of the D-instanton background.  
This transformation  is  
 different from the one considered in the first reference
 in \bbb, where S-duality was interpreted as a symmetry of the `rotated' action $( \del \p)^2 - e^{2\p}( \del \C)^2$. 
The basic $SL(2,R)$ transformation there was 
$S_{\pm } \to - 1/S_\pm$, \ $S_{\pm} = \C \pm e^{-\p}$, 
while \dfd\ corresponds in this notation  to $S_{\pm } \to  1/S_\pm$.
Both transformations are formally the symmetries of the scalar 
$SL(2,R)/SO(1,1) $  action
$-{\del S_+ \del S_-\ov (S_+ - S_-)^2}$  (see also \hull), but 
it is \dfd\   that is the standard 
$SL(2,Z)$ symmetry  that  has a  D-brane interpretation. 
More general $SL(2,Z)$ transformations \dfd\ do not 
admit a continuation to  the case of  imaginary $C$ which 
preserves reality
of the dilaton. 
Note that under the `Euclidean' transformation 
$S_{\pm } \to - 1/S_\pm$  one finds that 
$e^{  \p'} = e^{\p}-2  =  { q \l\ov u^4} -1$, 
i.e. that the  dilaton is defined 
in the `dual' region $u < u_*$ but it again blows up at $u=0$.}
\eqn\duu{
e^{  \p'} =2 - e^{\p} =  1 - { q \l\ov u^4} \ , 
\ \ \ \ \ \ \ \
\C' =  e^{ -  \p'}  -1  \ ,  } 
with the (unchanged) Einstein-frame metric
 again given by $AdS_5 \times S^5$.
 Thus  indeed the basic 
 S-duality transformation  corresponds to changing the sign of $q$ 
(and inverting 
the string coupling). 
The selfduality of D3-brane action  suggests
that switching of the sign of $q$ 
should be  induced by the   duality transformation 
on the SYM side.

The S-dual string-frame metric is  proportional to 
   \nit\ with $q \to -q$, 
\eqn\sdme{
(ds^2_{10})' = g_s^{-1} \a' \l^{1/2}  (1  - { q \l \ov u^4})^{1/2} 
\bigg( \l^{-1} { u^2}   dx_m dx_m +  {du^2 \ov u^2}  + d\Omega_5^2 \bigg)\  ,}
with   $u = u_*=(q\l)^{1/4}$ 
being  a  curvature  singularity.
The dual effective  string coupling in \duu\ 
remains {\it  small} for all values 
of $u > u_*$ and vanishes  at $u=u_*$.
Thus  under S-duality 
the  singularity  of the original background in the 
dilaton \diim,\hii\ (at $u=0$)  gets  `transformed' into the 
singularity in  the metric (at $u=u_*$).

The potential between monopole $m$ and anti-monopole $\bar{m}$ can be found 
by minimizing the
D-string action \dre.
 As  in the case  considered  in
\refs{\mono,\gross}, there exists a critical distance 
\eqn\yrt{
L_{\rm crit} \sim ({\l \ov q})^{1/ 4}\  , 
}
such that for  $L < L_{\rm crit}$, there is minimal D-string world sheet 
connecting $m$ and $\bar{m}$. For  $L \ll L_{\rm crit}$ 
we are effectively in the $AdS_5 \times S^5$ region 
and the potential 
is proportional to  $ 1/L$
as in the case of the  conformal
 $\N=4$ SYM theory.  
For  $L > L_{\rm crit}$ there is no string 
solution $u(\s)$
which minimizes \dre, i.e.   
the  potential becomes constant -- the  
 magnetic monopoles  are screened.
 Since the  D-string action \dre\ may be interpreted as an 
action of the  fundamental string 
propagating in the S-dual background  \sdme, 
the screening  behavior of the $m \bar{m}$ potential 
may  be understood also  from the fact that  around 
the `horizon'(singularity)  $u =u_*$  in \sdme\ 
 the string world sheet is 
contractable and can split into two separate parts.
 
 The resulting picture  is 
consistent with the   discussion in section 5:
 in the  region $L \geq  ({\l \ov q})^{1 / 4}$ 
 the  electric charges (W-bosons)
are  confined, while the magnetic monopoles are screened.

\newsec{Concluding remarks}
Our motivation
in discussing the above D3-brane + D-instanton 
background is that it provides a simple 
generalization  of the  gravity -- gauge theory 
correspondence in the pure D3-brane  case. 
We have seen that the  supergravity or string theory  description 
predicts a (coupling-independent) confinement behaviour, 
essentially because the  constant  D-instanton density
makes the string-frame metric  flat 
  in  the IR region.  This confinement does not apply to all states, 
and  there are still some  gapless excitations
in this background
(in particular, one linear combination of the dilaton and RR scalar
fluctuations satisfies the massless Laplace equation 
in $AdS_5 \times S^5$, see Appendix).

This  example  is different from 
the case of localized D-instantons 
in $AdS_5 \times S^5$   which correspond to localized
YM instantons  in the boundary gauge theory \refs{\adsi,\bala}
and are  virtual 
Euclidean configurations contributing to path integral 
and  producing non-perturbative contributions
to observables  on both sides of the  duality correspondence. 
Adding a   localized  instanton   
is  a  perturbation 
of the pure  D3-brane case, 
while  we  interpreted  the homogeneous self-dual gauge field 
  background 
as a special   $\N=2$ supersymmetric state of the $\N=4$ SYM  theory.

Though this  Euclidean self-dual background 
is  somewhat  artificial from physical point of view, 
it would  be interesting  to understand the reason 
for a power-like  IR flow of the gauge coupling \ruu\ 
 from the gauge theory point of view.

The conformal invariance of the $\N=4$ SYM  theory
is broken by the background  field  density  parameter $q$ in a `soft' way.
Since the Einstein frame metric remains that of the 
$AdS_5 \times S^5$  space, 
some of the  supergravity modes
which do not directly couple to the dilaton (like 
the $S^5$-radius mode) do not feel the presence of
 $q$, i.e.  still satisfy the  same  massive 
$AdS_5$ Laplace equations
as in the pure D3-brane case. 
The correlators of the corresponding gauge field operators
($F^4+...$ in the $S^5$ radius case) are thus expected to be the same as in the vacuum 
$\N=4$ SYM case. 
 At the same time, 
the dilaton  and RR scalar modes   which  satisfied the massless Laplace 
equation in the pure D3-brane case  no longer
do so in the D3+D(-1) background
(the correlators of $F^2$ and $FF^*$ operators are certainly different from the vacuum case).
Since the scalar field 
  background values are non-constant, their fluctuations
mix with each other and {\it also}
with  graviton perturbations.
The corresponding  quadratic fluctuation action is derived in Appendix below.
Such mixing  is   a general feature
of solutions with non-constant dilaton backgrounds, like 
D4-brane \refs{\haoz, \oog}   or the  solution of \refs{\kes,\gub,\min}.

\newsec{Acknowledgements}
We are   grateful to   M. Douglas, N. Itzhaki, I. Klebanov, J. Maldacena, 
J. Minahan, 
 M. Petrini  and  M. Shifman  
for  useful  discussions and comments 
at different   stages of this work.
This work was supported  by
 PPARC and 
 the European
Commission TMR programme ERBFMRX-CT96-0045, 
and  the INTAS grant No.96-538.

\vfill\eject
\appendix{A}{Perturbations  of   the  supergravity background}
Below we shall study small fluctuations
of the  metric, dilaton and RR scalar fields 
near the background  \diim,\hii,\nit.  
The equations for the dilaton and RR scalar perturbations are in general 
of interest in connection
with  spectrum of possible bound states
  on gauge theory side
\refs{\witten,\oog,\haoz}. The scalar perturbations are not,  in general, 
expected to decouple from the metric perturbations  in the  case  of 
  non-constant  scalar background
as we illustrate below on the present D3+D(-1) example.

We start with  the $D=5$  Einstein-frame 
Lagrangian  for the 
 metric-dilaton-axion system ($\C=i C$; $\m,\n=1,...,10$)
\eqn\oo{
L =  \sqrt{\hat{g}} \big[ \hat{R} - 
\ha  \, \hat{g}^{\m \n} (\del_{\m} \hat{\phi} 
\del_{\n} \hat{\phi} - e^{2 \hat{\phi}} \del_{\m} \hat{\C} \del_{\n} \hat{\C}) \big]
\ ,  } 
and expand the fields near their background values
\eqn\sool{
\hat{\phi} = \phi + \e\ , \,\,\,\,\,\, \ \ \ \ \ 
\hat{\C} = \C + e^{-\p} \x \ ,\,\,\,\,\,\,\ \ \ \ \
 \hat{g}_{\m \n} = g_{\m \n} + h_{\m \n}
\ . } 
The background fields $\phi$ and $\C$ satisfy
\eqn\dilaback{
\partial_\m \C =-  e^{- \phi} \partial_\m \phi\ , \ \ \   \,\,\,\,\,\,  \ \ \ \ 
 e^{-\p} D^2 e^\p = D^2 \phi + (\partial_\m \phi)^2   = 0\ , 
}
where $D_\m$   is the covariant derivative depending 
on the  background  metric $g_{\m \n}$. The  Einstein-frame 
metric in \nit\  is  that of the $AdS_5 \times S^5$ space. 

Expanding  to the second order in fluctuation fields we get  for the 
$\e$ and $\x$ dependent part  of $L$ (we  omit the pure graviton  $h^2$  part)
 \eqn\eeq{
L_2 = \sqrt{g}\bigg[  \td{h}^{\m \n} \del_{\m} \p [  \del_{\n} (\e + \x) -
 \del_{\n} \p\  (\e + \x)]
 - (\partial_\m \e)^2 + 2 (\partial_\m \phi)^2 \e^{2}
+  (\del_\m \x)^2  +  4 \del^\m \p \del_\m \e\  \x \bigg]
\ , }
where 
$$
\td{h}^{\m \n} \equiv  h^{\m \n} - \ha g^{\m \n} h, 
\,\,\,\,\,\,\,\,\,\,\,\, \ \ \ 
h = g_{\m \n} h^{\m \n} \ . 
$$
Following \hof\ we  choose  the diffeomorphism gauge fixing term to be 
$$ L_{g.f.} = - \ha \sqrt{g} \, g^{\m \n} 
[D_{\b} \td h_{\m}^{\b}  - \del_{\m} \p\ 
(\e + \x)] [D_{\a}\td  h_{\n}^{\a}  - \del_{\n} \p\ 
(\e + \x)]  
$$
\eqn\gah{
= \sqrt g \bigg[ -\ha  g^{\m \n} 
D_{\a} \td h_{\m}^{\a} D_{\b}\td  h_{\n}^{\b} 
+  D_{\n} h^{\m \n} \del_{\m} \p\  (\e+\x)
- \ha  (\del \p)^2  (\e +\x)^2 \bigg]
 \ . } 
The  $\e,\x$ dependent part of  the gauge-fixed Lagragian  is then 
(we again omit $ O(h^2) $ terms)
$$
L _2'= L_2 + L_{g.f.} = - \sqrt{g}\bigg[  \,  (\del_{\m} \p
\del_{\n} \p + D_{\m} D_{\n} \p)\ \td{h}^{\m \n} \ (\e + \x)
$$
\eqn\gaufl{
+ \  (\partial_\m \e)^2 - 2 (\partial_\m \phi)^2 \e^{2}
-  (\del_\m \x)^2  -   4 \del^\m \p \del_\m \e\  \x \ +  \ha (\del_\m \p)^2\ (\e + \x)^2\bigg] \ . 
}
Note that because of  the classical equation for $\p$ \dilaback\ 
the trace of $h_{\m\n}$  does not actually  couple to  $\e$ and $ \x$.

Using the explicit form of the 
background  Einstein-frame  metric and the dilaton  (see \diim,\hii,\nit)
\eqn\eex{
ds^2_{E} = {1 \ov z^2} ( dx_m dx_m + dz^2 + z^2 d \Omega^5)
\ , \  \ \ \ \ \ \ \ \ 
e^{\p} = 1 + {q z^4 \ov \l}
\ , } 
we get for the coupling  function in the  graviton -- $(\e+\x)$ 
mixing term 
\eqn\gett{
\del_{\m} \p \del_{\n} \p + D_{\m} D_{\n} \p = 
\d_{\m z} \d_{\n z} [\del_{z}^2 \p + {  z\inv} \del_z \p 
+ (\del_z \p)^2 ] - { z\inv} \d_{\m m} \d_{\n n} \del_z \p
\ .}
This function is non-vanishing, so 
  there is a non-trivial mixing between $\e+\x$  and the  
traceless part of $h_{\m \n}$. 

Introducing   
$$\e_{\pm} = \e \pm \x \ , $$
we may rewrite  \gaufl\  as 
\eqn\wre{
L _2' = - \sqrt{g} \, \bigg[ \del^\m \e_+ \del_\m \e_- 
- \ha (\del_\m \p)^2 \e_+^2 - 2 \del^\m \p \del_\m  \e_+ \ \e_- 
\ + \  (\del_{\m} \p \del_{\n} \p + D_{\m} D_{\n} \p)\  {h}^{\m \n}\  \e_+ \ \bigg] 
\ . }
Then the equations of motion for $\e_-$ and $\e_+$ are  
\eqn\moeu{\eqalign{
& D^2 \e_+ - 2 \del^\m \p \del_\m \e_+ -2 D^2 \! \p \, \e_+ = 0 \ ,  \cr
& D^2 \e_-  + 2 \del^\m \p \del_\m \e_- + (\del_\m \p)^2 \e_+  
+  (\del_{\m} \p \del_{\n} \p + D_{\m} D_{\n} \p) {h}^{\m \n} = 0 \ . 
}}
Using \dilaback\ the  equation  for  
 $\e_+$  can be written also as
\eqn\tyre{
D^2 (e^{-\p} \e_+) = 0
\ ,  } 
i.e. the field $ e^{-\p} \e_+$ satisfies the massless 
scalar equation in the $AdS_5 \times S^5$ space.

Since the   fields $\p$ and $\C$  couple  to the 
boundary gauge theory operators as  
  \eqn\ccq{ 
L_{int}(\p,\C) = e^{- \p} \tr (F_{mn} F_{mn}) -   \C \tr ( F_{mn} F^*_{mn})  \ , }
their perturbations couple as  
$$ 
L_{int}(\p+ \e ,\C + e^{-\p} \xi) = 
L_{int}(\p,\C)  - e^{-\p} \big[ \  \e \ \tr (F_{mn} F_{mn}) 
 +  \ \xi\  \tr (F_{mn} F^*_{mn}) \big]   + ...
$$ 
\eqn\cou{
= L_{int}(\p,\C)\   -\    e^{-\p}\e_+ \  \tr (F^{(+)}_{mn}F^{(+)}_{mn}) 
  -\   e^{-\p}\e_-\   \tr( F^{(-)}_{mn}F^{(-)}_{mn}  )
 + ...\ , } 
where
$$ F^{(\pm)}_{mn} \equiv \ha ( F_{mn} \pm F^*_{mn})  \ . $$
In interpreting this expression we should take into account
that $\e_+$ and $\e_-$  are
conjugate to each other  like  light-cone variables
(cf. the  kinetic term in \wre). Thus 
the term that couples to $\e_+$ is actually the source 
for $\e_-$ and vice versa.
In view of \tyre\ we then conclude  
that the operator 
$\tr (F^{(-)}_{mn}F^{(-)}_{mn})$  is 
still  counterpart of a 
massless mode in the $AdS_5$ space.
This is perfectly consistent with 
the fact that  the background  gauge field 
corresponding to the supergravity solution 
satisfies \den,  $< \tr(F^{(-)}_{mn}F^{(-)}_{mn})>=0,$ 
i.e. does not have a  condensate.

The second field $\e_-$ is  non-trivially coupled to the 
graviton  and  thus  has more  complicated dynamics, 
reflecting  the fact that  the operator 
$\tr (F^{(+)}_{mn}F^{(+)}_{mn})$ 
should `feel' the presence of the gauge field background.

\baselineskip 10pt plus 2pt minus 2pt

\vfill\eject
\listrefs
\end